# Study of the coherent perturbation of a Michelson interferometer due to the return from a scattering surface

Vitalii Khodnevych, Sibilla Di Pace, Jean-Yves Vinet, Nicoleta Dinu-Jaeger and Michel Lintz
ARTEMIS, Universite and Observatoire de la Cote d'Azur, CNRS, Nice, France

## ABSTRACT

We describe a setup based on Michelson interferometry for coherent measurements of the backscattered light from a low roughness optical surface under test. Special data processing was developed for the extraction of the useful signal from the various stray contributions to the coherent signal. We achieve coherent detection of light scattered by a mirror down to -130 dB in optical power. We characterize the dependence of the backscattered light with spot position and incidence angle. Results of cross-polarization scattering coherent measurements and preliminary results of dust deposition experiment are presented here. This work represents the first step in the experimental evaluation of the coherent perturbation induced by the scattered light in the space gravitational wave detector of the LISA mission.

## 1. INTRODUCTION

The Laser Interferometer Space Antenna (LISA) is a space-based gravitational wave observatory [1] now in Phase A. It is represented by a triangular constellation of three identical satellites, each satellite containing two gravitational reference sensors (e.g. test masses) at the end of each arm. The measurement of length variation between the two test-masses, variation driven by gravitational waves but also by various noise sources, is provided by heterodyne interferometry.

Any source of coherent scattered light can perturb the interferometric measurement since it can give rise to a significant noise during the heterodyne phase measurement at a level of one or several micro-radians.

Before evolving strategies to mitigate scattered light in the LISA instrument, and since the backscattered light in the telescope can affect the long arm (2.5 millions of km) length measurement, a good understanding of the perturbation of heterodyne interferometry by scattered light has to be achieved prior to mission launch.

## 2. SETUP

As a first step, we have implemented a setup for the coherent measurement of backscattered light (see fig. l). The setup is split in two parts: a fibered one, containing the laser source and a fibered Michelson interferometer, and a free space one with optics designed to illuminate the sample under test with a collimated beam and to collect the backscattered light. A lock-in detection is used to drive a phase modulation applied to the beam and to demodulate the fringe signals.

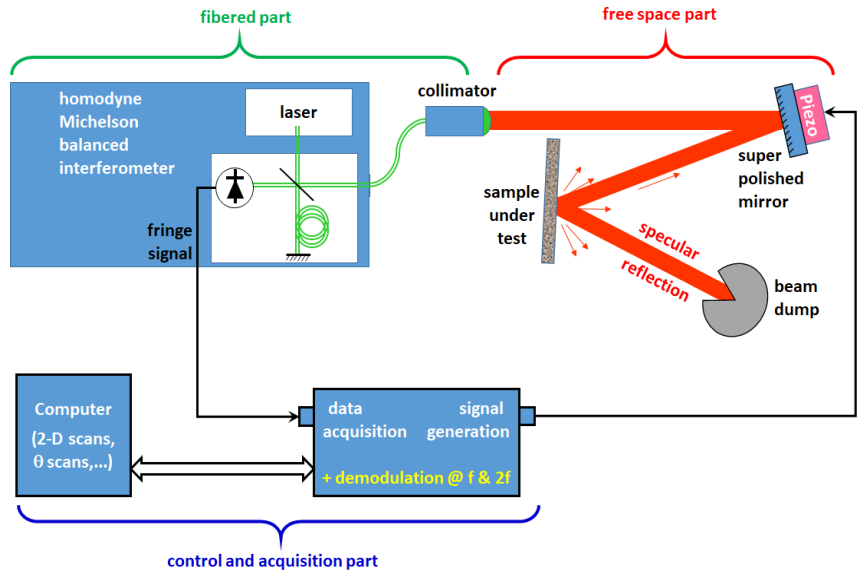

Figure 1. Block diagram of the coherent back-scattering measurement set-up

The detailed schematics of the setup is shown in figure 2. The basis of this setup is a Fibered Michelson interferometer. We use a narrow linewidth Orion RIO infrared laser diode of 1.542 µm wavelength and 10.2 mW power. An optical circulator, inserted between the laser source and the beam splitter, prevents the laser from receiving the fraction of the light which is rejected by the Michelson interferometer and could otherwise cause amplitude and frequency modulation in the laser. Instead, the rejected beam is sent to a second photodiode (PD 2), to increase the gain and precision of the measurements. Then the light is divided by the fibered 50/50 beam splitter and follows the two arms of the interferometer. Arm I is fully fibered, with a fibered 100% reflector and, if a scan of the optical phase difference is useful, a thermal ramp can be applied. Arm II is partially fibered, up to a collimator, which emits a collimated beam for the free space part. All optical fibers are polarization maintaining.

The collimated beam in the free space propagates to a super polished mirror with a piezo actuator for rapid modulation of the optical phase. The piezo is powered by a 2 kHz sine from the internal signal generator of the lock-in amplifier. From the piezo mirror, the collimated, phase modulated beam propagates to the sample under test, its backscattering being the aim of our measurements. The backscattered light which is recoupled into the fiber interferes with the beam from Arm I. A beam dump [2] is placed in such way to attenuate efficiently the specularly reflected beam from the sample. In the presence of the backscattering, the interference between ARM I and ARM II beams gives rise to a modulation in the photodiode signals. The modulated signal are amplified, and then demodulated by a lock-in amplifier.

As will be shown in the following, the scattered light has a speckle behavior as a function of the incidence angle, and it is possible to adjust the position and the incident angle at the beam dump to minimize its backscattering. The beam dump is made from two black HOYA RT-830 glass plates polished by Coastline Optics. The two absorbing glass plates are placed, parallel to each other, so that the lasers beam bounces a number of times and is attenuated. The minimized backscattering, which we measure from this beam dump is -130.9±0.9 dB in optical power. To separate the contribution of the beam dump from other contributions to backscattering, we modulate the position of the beam dump at the frequency of 1 Hz, and identify the corresponding contribution in the fringe signal. When the incidence at the beam dump is well adjusted, we switch the 1 Hz signal to modulate the sample position. With properly adjusted amplitude we completely discriminate the contribution of the backscattering of the sample against other contributions (see below). To make 2D maps of the backscattering amplitude, we use a 2D translation stage controlled from a PC to move the sample parallel to its surface. For angular study of scattered light, we use a Picomotor rotational stage (New Focus 8821) instead of the 2D stage.

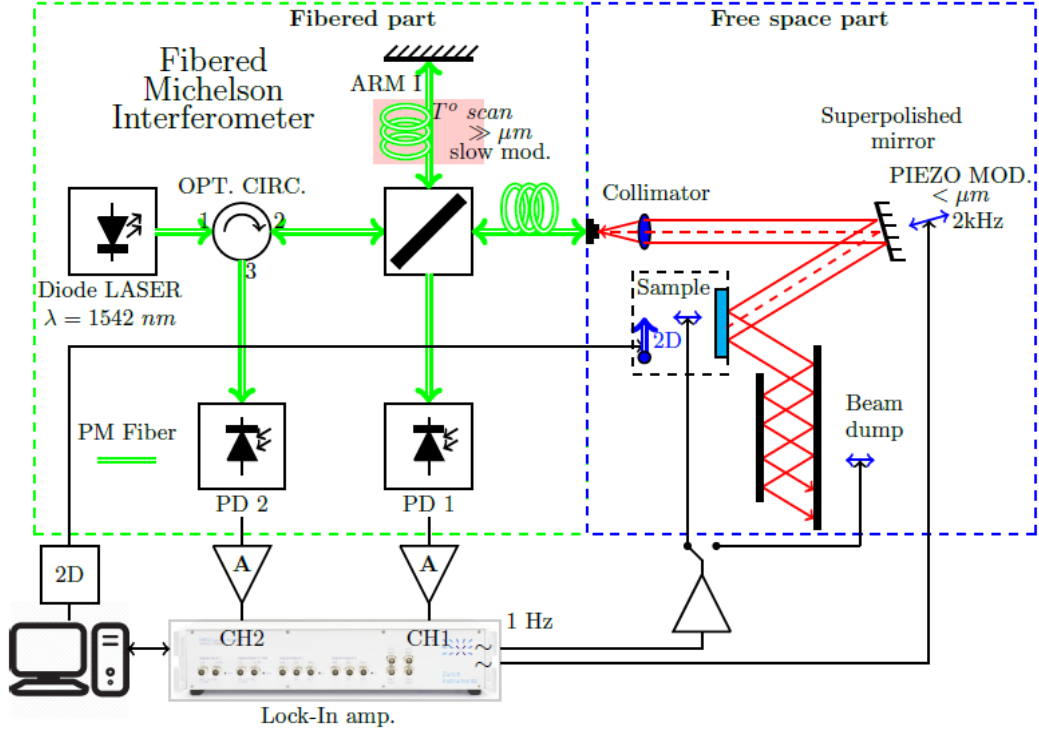

Figure 2. Coherent backscattering measurement setup. Green: optical fibers. Red: free space beams. Dashed red: backscattered light from the sample.

The same interferometric set-up is used for the measurement of coherent backscattering from dust deposition (see Results section and Fig. 9). A controlled deposition of a given size and number of dust particles on the mirror is obtained by using a spray system. A camera and a white light illuminating LED are placed in front of the sample. Using image processing of the camera images, the information related to the number of particles on the sample and it is subsequently correlated with the measured interferometric signal.

## 3. DATA PROCESSING

Ideally, after recombination at the 50/50 beam splitter, the backscattered fraction of light from the sample surface, with an amplitude of $A = \sqrt{\frac{I_L}{4}} \sqrt{b_s} \exp(i\phi_{II,s})$, interferes, with the light of amplitude $A = \sqrt{\frac{I_L}{4}} \exp(i\phi_I)$. At the PD 1 (resp. PD 2) output ports of the interferometer, the interference signal is $I = \frac{I_L}{2}\sqrt{b_s}\cos(\Delta\phi_s)$ (resp. is $-\frac{I_L}{2}\sqrt{b_s}\cos(\Delta\phi_s)$). In this expression $I_L$ is the laser intensity, $\Delta\phi_s = \phi_{II,s} - \phi_I$ is the difference of the propagation phase between arm II (up to the sample), including the 2 kHz modulation of the piezo actuator, and the arm I, including a possible thermal ramp. Then extracting the backscattered power fraction $b_s$ by demodulation looks trivial.

However, the processing of demodulated signals is complicated, because multiple contributions from different backscattering sources (e.g. fibers, fiber connectors, collimator, piezo mirror, beam dump, and sample) are presented in the measured signal. First, one could consider separately the three sources located before the piezo mirror, from which backscattering signal should not be modulated, and discard them. But the acoustic crosstalk between the fibered part and the piezo actuator is not negligible, as we clearly observe the corresponding contribution in the demodulated signal. Other contributions to backscattering come from the modulated mirror and from the beam dump. Therefore, the total fringe signal at PD 1 has to be written as following:

$$I = \frac{I_L}{2}\left(\sum_i \sqrt{b_i}\cos(\Delta\phi_i) + \sqrt{b_s}\cos(\Delta\phi_s)\right) \qquad (1)$$

where all stray contributions to backscattering (power ratio $b_i$ and phase $\phi_i$) are under the sum sign. Demodulation is done at the first harmonic (frequency 2 kHz) with a bandwidth of 6 Hz (8th order). After demodulation of the signal using the lock-in amplifier, we get 224 times per second, in-phase X and quadrature Y components. The combination we need is:

$$R \equiv \cos(\Theta) \times X + \sin(\Theta) \times Y \qquad (2)$$

where $\Theta$ is the phase of the fringe signal modulation with respect to the voltage applied to the piezo actuator. It differs from zero due partly to the electronic signal propagation delay and mainly to the hysteresis in the piezo actuator. As $\Theta$ does not change during data taking we first determine the value of $\Theta$ in conditions where the signal is large. Later we use this value to determine R from eq.2, and exploit the time series of R, particularly the dependance of R on the thermal drift of the fibered path length and/or on the modulations we apply to the position of the sample.

Then the demodulated signal of the measured signal at the first and second harmonics of piezo mirror modulation frequency (2 kHz) will be:

$$R_1 = I_L \left( \sum_i \sqrt{b_i} J_1(2\pi\delta_{PMi})\cos(\phi_i) + \sqrt{b_s} J_1(2\pi\delta_{PM})\cos(\phi_s) \right)$$

$$R_2 = I_L \left( \sum_i \sqrt{b_i} J_2(2\pi\delta_{PMi})\sin(\phi_i) + \sqrt{b_s} J_2(2\pi\delta_{PM})\sin(\phi_s) \right) \qquad (3)$$

where $J_1$ is Bessel function of first order and $\delta_{PM}$ is the modulation depth of the piezo mirror (note that for each component modulation depth can be different). Separation of the contributions from all backscattering sources is a difficult and unnecessary task. What we need is to give a specific signature to the $\sqrt{b_s}$ contribution, to identify and record it with the highest precision. For this, we modulate the position of the sample at $\frac{\omega_s}{2\pi}$ =1 Hz. Using Fast Fourier Transform (FFT) we can easily extract all harmonics of the modulated signal from $R_1$. When modulation of the sample is applied, we have an additional term in phase $\phi_s$ for sample and beam dump components:

$$R_1 = I_L \left( \sum_i \sqrt{b_i} J_1(2\pi\delta_{PM})\cos(\phi_i) + \sqrt{b_s} J_1(2\pi\delta_{PM})\cos(\phi_s + \delta_1 cos(\omega_s t)) \right.$$
$$\left. + \sqrt{b_{BD}} J_1(2\pi\delta_{PM})\cos(\phi_{BD} + 2\delta_1 cos(\omega_s t + \phi_0)) \right) \qquad (4)$$

Similar for the second harmonic. Here $\delta_1$ is the slow frequency modulation depth and $\phi_0 \approx 0$ is a propagation delay (sample - beam dump – sample). The factor two in front of the modulation term considers that the backscattered light from the beam dump is twice modulated. In fig. 3 is presented the FFT of $R_1$ in different measurement conditions, but for the same scattering point. In red is the FFT of $R_1$ when 1 Hz modulation is applied to the sample. At very low frequency (below 1 Hz) are spectral components of the parasite backscattering contributions. They are not modulated, so they contribute only on frequencies <1 Hz, due to thermal drifts. At higher frequencies modulation harmonics at 1 Hz of useful signal mixed with the beam dump backscattering contribution. It's interesting to recall that functions of type $\cos(\phi + \delta\sin(\omega_s t))$ can be presented as series:

$$J_0(\delta)\cos(\phi) + 2\sum_n J_{2n}(\delta) \cos(2n\omega_s t)\cos(\phi) - 2\sum_n J_{2n-1}(\delta) \cos((2n-1)\omega_s t)\sin(\phi) \qquad (5)$$

So, the distribution of the power between harmonics is given by the Bessel function of the first kind. Argument of this function is the modulation depth $\delta$. By adjusting the $\delta$, we can redistribute the power in the spectrum in a desired way. This property is useful, as a spectrum (5) has slow (thermal) component multiplied on $J_0$. If we adjust $\delta$, so that $J_0(\delta) = 0$, then we will have useful signal separated in frequency from parasite ones. So, for this procedure, it is crucial to adjust the modulation depth of the sample in a way to reach the zero of Bessel function of zero order.

At this stage, we had separated the useful signal from the parasite contribution. But it's still mixed with backscattering from the beam dump. To lower the contribution from the beam dump, we adjust its position, to reach the minimum possible scattering. This is possible because scattered light has speckle behavior (as will be shown later). The fig. 3 shows the FFT of $R_1$ when slow modulation is applied to the beam dump and not to the sample (black color). The contribution from the beam dump is approximately one order of magnitude lower than the backscattering amplitude measured from the sample. So, this contribution can be neglected.

The Figure 3 shows also the FFT of $R_1$ when slow modulation is not applied (green color). In this case, the contributions of backscattering are not separable, and any precise measurements are not possible. In blue in the same figure is shown the case when neither slow nor fast modulation is applied to the setup. So, this displays the contributions to the noise floor from the laser power noise and electronic noise of the detection system.

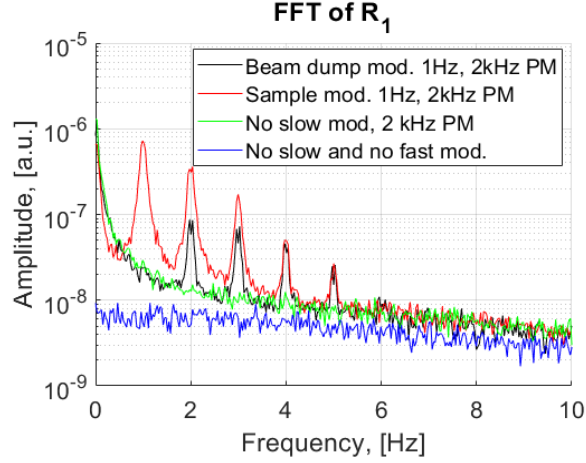

Figure 3. FFT of $R_1$ when slow (1 Hz) modulation is applied to the beam dump (black), and the case without slow modulation (green). For the sample modulation (red) we had adjusted the voltage to reach the condition $J_0(\delta) = 0$. The case when neither the fast, nor the slow is applied gives electronics plus laser noise level (blue).

After all this adjustments and assumptions, we will have:
$$R_1 = I_L\sqrt{b_s}J_1(2\pi\delta_{PM})\cos(\phi_s)$$
$$R_2 = I_L\sqrt{b_s}J_2(2\pi\delta_{PM})\sin(\phi_s) \qquad (6)$$

Recalling that $[\sin(\phi_s)]^2 + [\cos(\phi_s)]^2 = 1$, the combination with the second harmonic contribution will allow getting rid of the impact of thermal variations of the phase $\phi_s$

$$\sqrt{b_s} = \sqrt{\left(\frac{R_1}{2J_1(2\pi\delta_{PM})}\right)^2 + \left(\frac{R_2}{2J_2(2\pi\delta_{PM})}\right)^2}\frac{1}{I_L/2} \qquad (7)$$

The calibration coefficient $I_L/2$ is measured when in the ARM II the sample is replaced by a metal mirror at normal incidence. To reduce statistical error, we combine the measured values from the two photodiodes (weighted mean).

In this section we have shown that, since the measured signal is contaminated by different backscattering sources, a specific data processing algorithm (eq. 7) has to be developed, which, with few adjustments in the set-up, allows extracting the useful data from the measured signal. The current method allows to measure the backscattering from the sample down to the level of ≈-130 dB in optical power, limited by the backscattering from the beam dump.

# 4. RESULTS

Speckle behavior of scattered light is observable when the sample is moved parallel to its optical surface, using the setup shown in Fig. 2. The correlation length under the translation of the sample is compatible with the beam waist (see fig. 4). In this figure is shown measured backscattering fraction in amplitude, when we translate the sample. The beam shape shown in black is a Gaussian intensity profile with the size of the collimated beam ($\omega_0$=1.71 mm).

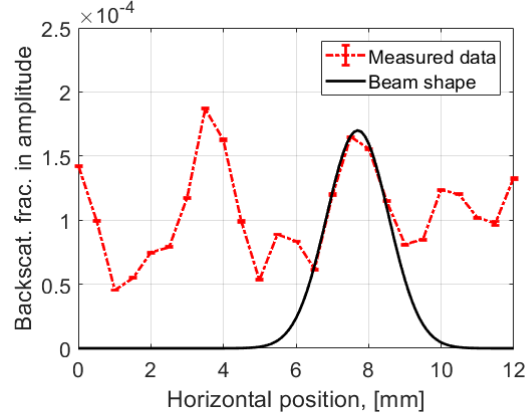

Figure 4. Profile of backscattered light from a moderately scattering target. Black curve (arbitrary units): plotted for comparison, a gaussian beam profile of the same gaussian beam size ($\omega_0$=1.71 mm) as the incident beam.

For the sample under test, a 1-inch "IR mirror" from Edmund Optics (ref. 47113) has been used. It has RMS roughness less than 175 Å. This value is large, however it is small enough that scattering is still in the "small roughness" regime. The mirror roughness is probed by a collimated beam with gaussian size $\omega_0$=1.71 mm obtained from an 18 mm focal length collimator. The 2D translation of the sample is carried out by the Newport translation stage with a step 0.5 mm over a distance of 12 mm. The beam in the free space part of the setup has a beam divergence angle of 0.58 mrad.

The 2D maps in Figure 5 show rapid changes with the incidence angle at the sample. Because the change of incidence is small compared to the beam divergence, one can follow the correlation between successive scattering maps.

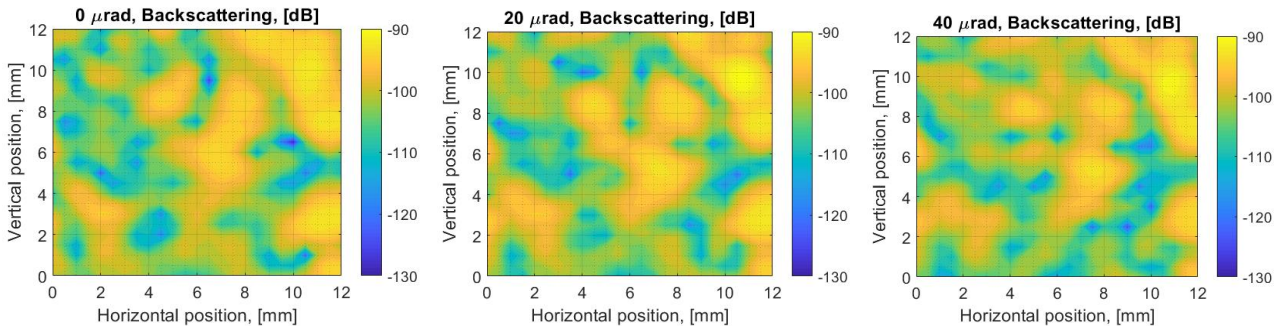

Figure 5. 2D maps of the backscattered power (in dB) of the same scanning area with slightly different incidence angle (20 µrad steps). Incidence angle is approximately 14°. Horizontal and vertical scan step is 0.5 mm. dB is for the fraction of optical power of the backscattered light recoupled into the fiber.

The ratio between maximum and minimum scattering points in each of these maps reaches 3 orders of magnitude in optical power. On the other hand, the RMS average value of the backscattering power for each map changes by a fraction of a dB (98.4, 98 and 98.6 dB respectively).

It is natural to expect, that after a change of the collimator to generate a different Gaussian beam waist, the map of scattered light will change, but the spatial speckle properties will remain similar up to the speckle grain size. For example, after replacing the collimator, the beam waist changes from $\omega_0$=1.71 mm to $\omega_0$=0.65 mm (and the divergence increases from 0.58 to 1.51 milliradian). Map measured with 0.25mm step size and 12x12 mm² scanning area is shown in figure 6 left. One row of this map with overlapped beam shape is shown in the figure 6 right.

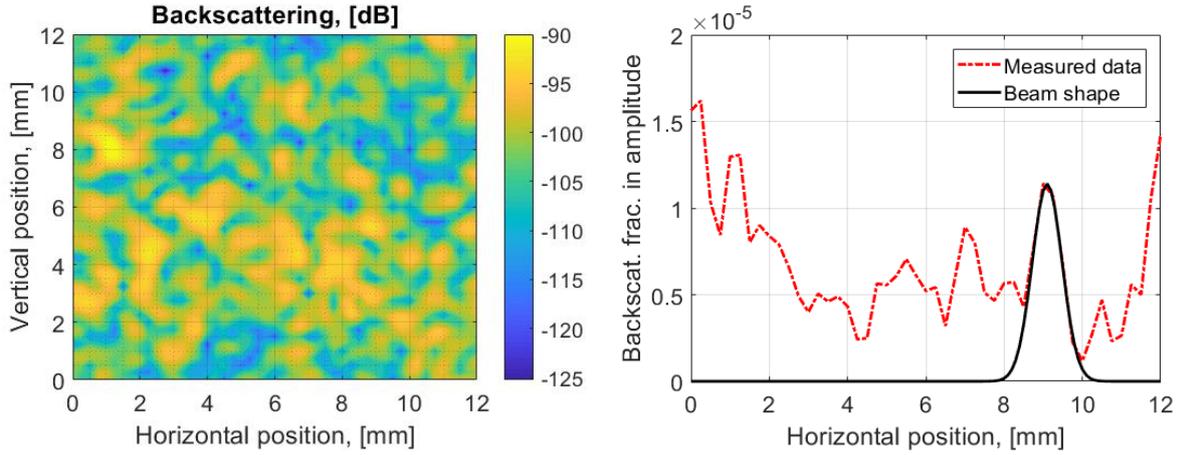

Figure 6. Map of backscattered light with a collimator of 8mm focal length. Size of scattering peak is same as Gaussian beam size ($\omega_0$=0.65 mm), in black.

The angular study was made with a Picomotors rotational setup with the step of 6.3 µrad over a range of 6.3 mrad (see fig. 7). On the same figure is present Gaussian of angular beam shape. Size of this Gaussian (size of speckle grain) is got from the first zero of the autocorrelation function and is $\Theta_0$=0.233 mrad, when the total divergence angle is 1.51 mrad. The angular beam shape is given by formula:

$$I = I_0 \times \exp(-2\Theta^2/\Theta_0^2) \tag{8}$$

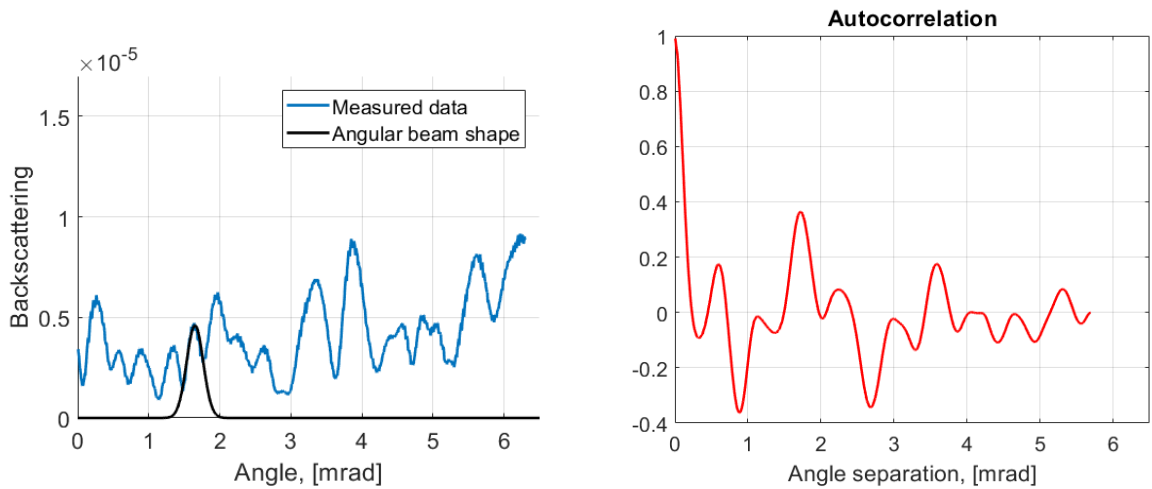

Figure 7. Left: Dependence of the backscattered light amplitude with the incidence angle. Waist of the collimated beam: 0.65mm. In black: gaussian beam divergence profile of the beam amplitude, with half-angle divergence angle 0.233 milliradian. Right: corresponding autocorrelation function of the backscattering amplitude.

In the LISA project, the "long-arm" heterodyne interferometer involves a polarizing beam splitter (PBS), to prevent stray light from the Transmitted beam to the far satellite to reach the interferometer. But this polarization-based rejection is limited by the rejection ratio of the PBS, as well as by the fraction of light which is back-scattered, for instance by the mirrors of the telescope, with crossed polarization. For that reason we recorded the backscattering coherent signal in a configuration where we insert a Faraday mirror, to rotate by 90° the polarization in Arm I. We used the same translation stage to record the coherent signal as a function of the transverse position of the sample. We made measurements both with vertical and horizontal incident polarization (see fig. 8). Backscattered light in crossed polarization has lower amplitude and does not strictly follow the backscattering profile in parallel polarization.

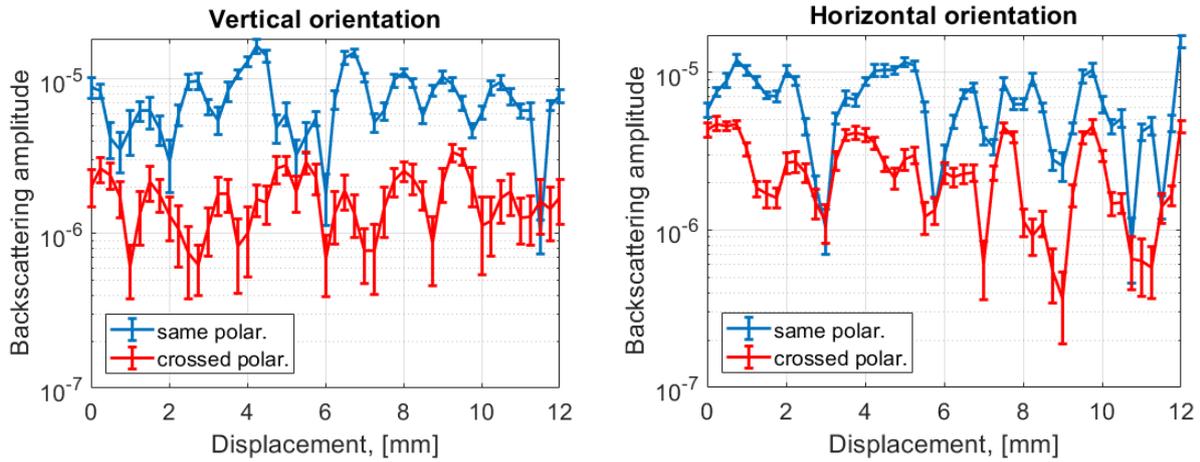

Figure 8. Measurements of backscattered light with crossed polarization, with vertical incident polarization (left) and horizontal incident polarization (right). Beam waist: 0.65mm. The scan step is 0.25 mm.

The amplitude of the backscattered light in the crossed polarization is not negligible. More detailed study will be done in this field in the future.

To estimate the dependence of the coherent backscattering amplitude on the number of particles on the optical surface, we have started preliminary measurements of the coherent backscattering from a mirror, initially clean, with dust deposition. The interferometric setup has the same structure as presented in fig. 2. To make consecutive pictures of the mirror, we have installed the camera in front of the mirror and an LED illumination system. Using an image processing algorithm we identify the number of particles at the surface, while using the interferometric setup, we measure the coherent backscattering. Dependence of the backscattering amplitude on the number of particles is shown in fig. 9.

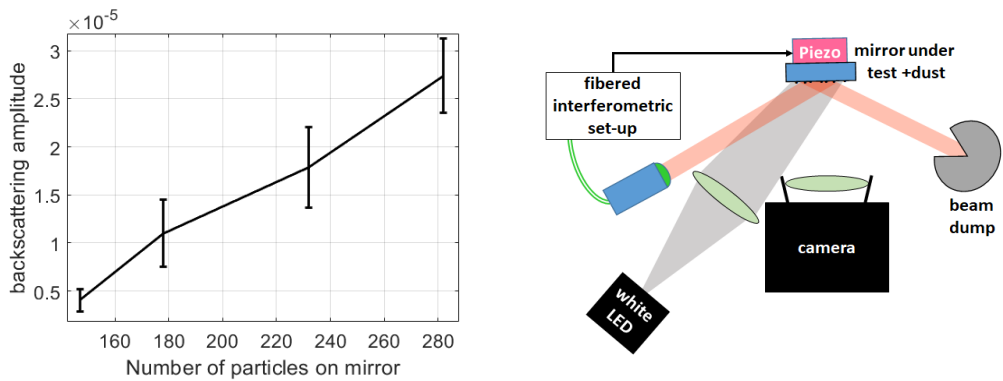

Figure 9. Left: the dependence of the fractional backscattering $\sqrt{b_s}$ on the number of particles at the surface (preliminary results). Gaussian size of the laser beam: 1.71 mm. Backscattering is measured in parallel polarization. Right: coherent backscattering measurement set-up for the dust-contaminated mirror.

## 5. CONCLUSIONS

We have developed a setup and data processing for the measurements of the backscattered light from an imperfect surface with a view at the scattered light issues in the LISA mission. We have studied scattered light from an IR mirror with intermediate roughness (less than 175 Å rms roughness). We have observed, in the coherent backscattering amplitude, speckle-type dependence with spot position and incidence angle, as can be expected from the stochastic nature of the roughness profile. We had found that size of speckle grain corresponds to the size of the beam. Crossed polarization of backscattered light is present and it is not negligible. These observations will be improved, particularly by measuring the dependence with large incidence angle changes, for comparison with models based on the measured roughness profile, and with models based on the measurement of the bi-directionnal reflectance distribution function (BRDF) of the samples.

Preliminary observation was made of the correlation between the coherent backscattering amplitude and the number of dust particles. As preventing the deposition of dust during the assembly and launch phases of the LISA mission will be difficult, this study will be improved and extended. White light images will be used to retrieve, not only the number of particles, but also their size and position, and to model the coherent scattering from the observed dust distribution.

## ACKNOWLEDGMENTS

This work is co-funded by: Region PACA, Thales Alenia Space, CNES, CNRS.

## REFERENCES

[1] LISA Mission Proposal for L3 submitted to ESA. June 2017
https://www.lisamission.org/?q=articles/lisa-mission/lisa-mission-proposal-l3

[2] M. Lintz, A.V. Papoyan, Review of Scientific Intruments 71 (2000) p. 4681, "A simple and efficient laser beam trap using a highly absorbing glass plate at Brewster incidence"